
\font\lbf=cmbx10 scaled\magstep2

\def\bs{\bigskip}
\def\ms{\medskip}
\def\np{\vfill\eject}

\def\ni{\noindent}
\def\cl{\centerline}

\def\title#1{\cl{\lbf #1}}
\def\ref#1#2#3#4{#1\ {\it#2\ }{\bf#3\ }#4\par}
\def\refb#1#2#3{#1\ {\it#2\ }#3\par}

\def\AP{Ann.\ Phys.}
\def\CQG{Class.\ Qu.\ Grav.}

\def\CPAM{Comm.\ Pure App.\ Math.}

\def\PR{Phys.\ Rev.}

\def\ZP{Z.\ Phys.}

\def\d{\hbox{d}}
\def\e{\hbox{e}}
\def\p{\partial}
\def\S{S_\circ}

\def\f#1#2{{\textstyle{#1\over#2}}}

\magnification=\magstep1

\title{Relativistic thermodynamics}
\bs\cl{\bf Sean A. Hayward}
\ms\cl{Yukawa Institute for Theoretical Physics, Kyoto University, 
Kyoto 606-8502, Japan}
\ms\cl{\tt hayward@yukawa.kyoto-u.ac.jp}
\bs\ni{\bf Abstract.}
A generally relativistic theory of thermodynamics is developed,
based on four main physical principles: heat is a local form of energy,
therefore described by a thermal energy tensor; conservation of mass, 
equivalent to conservation of heat, or the local first law; 
entropy is a local current; 
and non-destruction of entropy, or the local second law.
A fluid is defined by the thermostatic energy tensor being isotropic. 
The entropy current is related to the other fields by certain equations,
including a generalised Gibbs equation for the thermostatic entropy, 
followed by linear and quadratic terms in the dissipative 
(thermal minus thermostatic) energy tensor. 
Then the second law suggests certain equations 
for the dissipative energy tensor,
generalising the Israel-Stewart dissipative relations,
which describe heat conduction and viscosity 
including relativistic effects and relaxation effects.
In the thermostatic case, the perfect-fluid model is recovered.
In the linear approximation for entropy, the Eckart theory is recovered.
In the quadratic approximation for entropy, 
the theory is similar to that of Israel \& Stewart,
but involving neither state-space differentials,
nor a non-equilibrium Gibbs equation, nor non-material frames.
Also, unlike conventional thermodynamics, 
the thermal energy density is not assumed to be purely thermostatic,
though this is derived in the linear approximation.
Otherwise, the theory reduces in the non-relativistic limit 
to the extended thermodynamics of irreversible processes due to M\"uller.
The dissipative energy density seems to be a new thermodynamical field, 
but also exists in relativistic kinetic theory of gases.
\bs\cl{PACS: 05.70.-a, 04.20.-q, 04.40.-b}
\bs\cl{Revised 6th January 1999}
\np\ni
{\bf I. Introduction}
\ms\ni
Thermodynamics is perhaps best known to many relativists 
by its analogies with black-hole dynamics,
made concrete by the famous result of Hawking that 
stationary black holes radiate quantum fields with a thermal spectrum.
Recently, another link has been obtained: 
a unified first law which contains first laws of both black-hole dynamics 
and relativistic thermodynamics,
as described in an accompanying paper [1].
This result requires little thermodynamics other than 
the general form of the energy tensor given long ago by Eckart [2].
However, the literature does not seem to contain 
a fully satisfactory general theory of relativistic thermodynamics, 
in the classical sense of a general macroscopic theory independent of 
(but consistent with) microscopic theories. 
This remarkable situation has prompted the development 
of the theory described in this paper.
Such a theory naturally has widespread applications 
in astrophysics and cosmology.
For simplicity, only a single material will be considered here,
ignoring mixtures, phase changes, chemical reactions and external forces.

Perhaps the greatest progress in relativistic thermodynamics 
was made by Eckart [2], who:
identified the basic fields, formulated in this paper as the material current, 
the thermal energy tensor, the entropy current and the temperature;
gave the basic conservation equations, of mass and energy-momentum;
related the entropy density for fluids by a relativistic Gibbs equation;
and gave a relativistic second law, showing that it could be satisfied 
by simple dissipative relations which generalise 
the Fourier and Navier-Stokes equations for heat conduction and viscosity.
Thus the thermodynamic system is described by a finite number 
of macroscopic fields which may be measured by standard techniques.

Although Eckart was concerned with special relativity,
the theory generalises immediately to general relativity.
The only unacceptable part of the Eckart theory is that 
the dissipative relations entail non-causal propagation of disturbances 
in temperature and velocity.
The problem of infinite propagation speeds 
existed even in non-relativistic thermodynamics for longer still,
until M\"uller [3] showed how to resolve it 
by allowing the entropy current to contain terms 
which are quadratic (rather than just linear) in the dissipative fields,
namely thermal flux and viscous stress.
Physically, this means assuming that the dissipative fields are small enough 
to justify the approximation: thermostatic, linear or quadratic.
This incidentally means that entropy is generally not a state function,
as claimed in classical thermodynamics,
though this is retained in the linear approximation.

One might expect to obtain an acceptable theory of relativistic thermodynamics 
simply by applying a M\"uller extension to the Eckart theory.
This was attempted by Israel \& Stewart [4],
who obtained modified dissipative relations 
which are consistent with the M\"uller theory.
However, they gave a type of Gibbs equation 
which differs from that of Eckart and does not reduce appropriately to 
the Gibbs equation of non-relativistic local thermodynamics,
as described for example by de~Groot \& Mazur [5].
Also, Israel \& Stewart gave a theory intended to hold 
not only in the material frame of Eckart, 
but also in the frame of Landau \& Lifshitz [6]. 
This seems to involve some further approximation which is unclear, 
at least to this author.
In any case, this paper will use the material frame only,
for reasons explained below.

The theory described in this article was developed from first principles,
but includes M\"uller-type quadratic dissipative effects,
without which the Eckart theory is recovered.
The theory is based on simple physical principles 
which are consistent with local non-relativistic thermodynamics, 
including local first and second laws.
These principles serve to define what is meant by thermodynamic matter,
which is intended to describe real fluids and solids 
in circumstances where a macroscopic description is adequate.
The most basic principle is that {\sl heat is a local form of energy}.
In relativity, heat is therefore described by an energy tensor,
called the thermal energy tensor,
whose various components are thermal energy density, 
thermal flux and thermal stress.
Thermal flux is what is normally called heat flux only in the material frame,
thereby fixing this frame as the physically natural one.
Thermal stress is what is normally called material stress.
Thermal energy is what is normally called internal energy, but with fixed zero.

This principle entails one further generalisation:
the thermal energy density need not be purely thermostatic,
allowing a further dissipative field, namely dissipative energy density.
Classical thermodynamics excludes this by the claim that 
internal energy is a state function.
However, this is inconsistent with relativistic kinetic theory of gases [7], 
where thermal energy density is defined in terms of the molecular distribution 
and is generally not purely thermostatic.
In fact, dissipative energy density emerged unrecognised 
even in non-relativistic kinetic theory 
as essentially the 14th moment of the 14-moment approximation of Grad [8].
It turns out that relativistic causality requires 
dissipative energy density to vanish in the linear approximation for entropy, 
thereby recovering the Eckart theory.
However, this need not hold in the quadratic approximation,
thereby generalising the Israel-Stewart dissipative relations.

This article is intended for relativists 
who may have no previous experience of thermodynamics.
(Thermodynamicists with a nodding acquaintance with relativity
should also find it accessible).
Therefore brief reviews are given of M\"uller's extended thermodynamics,
of the previously standard local (or field) thermodynamics,
and of relevant classical thermodynamics.
This seems necessary because most textbooks on thermodynamics, even today,
insist on applying thermostatic concepts 
in ways which are quite inapplicable in genuine thermodynamics,
even in the formulation of the first and second laws.
Such confusions often survive in work on relativistic thermodynamics.

The article is organised as follows.
Sections~II--VI respectively review classical thermodynamics,
non-relativistic hydrodynamics, non-relativistic thermodynamics,
extended thermodynamics and basic general relativity.
Section~VII introduces the relativistic concept of heat and the local first law,
or conservation of heat, equivalent to conservation of mass.
Section~VIII checks the Newtonian limit.
Section~IX divides the thermal energy tensor 
into thermostatic and dissipative parts 
and defines a fluid by the thermostatic energy tensor being isotropic.
Section~X introduces the relativistic concept of entropy 
and the local second law,
and describes the thermostatic case using a generalised Gibbs equation.
Section~XI describes the linear approximation for entropy,
showing that it reduces to a minor generalisation of the Eckart theory.
Section~XII describes the quadratic approximation for entropy,
obtaining dissipative relations which generalise the Israel-Stewart relations
(in the material frame).
Section~XIII concerns integral quantities and corresponding laws.
Section~XIV concludes.
A point of style is that 
equations in the text are sometimes used for temporary arguments,
whereas displayed equations are intended to carry more weight.
\np\ni
{\bf II. Classical thermodynamics}
\ms\ni
In classical thermodynamics, the basic quantities are temperature $\vartheta$,
heat supply $Q$, work $W$, internal energy $H$ and entropy $S$.
The classical first law is said to be $\d H=\delta Q+\delta W$
and the second law $\vartheta\d S\ge\delta Q$,
where $\d$ is said to be a differential and $\delta$ not,
neither being actually defined except in thermostatics,
where the second law becomes an equality, or in the limit where they become 
difference operators between two equilibrium states separated in time,
corresponding to the classical experimental situation.
Clearly this is nonsense,
as has been emphasised eloquently by Truesdell [9].
The resolution is simply that 
these operators are really derivatives with respect to time.
In fluid mechanics the relevant derivative is that along the fluid flow, 
called the material or comoving derivative, henceforth denoted by a dot.
Similarly, in solid mechanics the relevant derivative is that
in the centre-of-momentum frame,
which may also be called the material derivative.
Then the first law is
$$\dot H=\dot Q+\dot W\eqno(2.1)$$
where $\dot Q$ and $\dot W$ should be taken as the basic quantities,
$Q$ and $W$ themselves not being uniquely defined.
For instance, for an inviscid fluid, the work is given by $\dot W=-p\dot V$,
where $V$ is the volume and $p$ the pressure of the fluid,
so that the first law reads
$$\dot H=\dot Q-p\dot V.\eqno(2.2)$$
Similarly, the second law is
$$\vartheta\dot S\ge\dot Q.\eqno(2.3)$$
Both laws now make sense, 
though it turns out that they respectively require 
the pressure and temperature to be uniform, 
again reflecting the classical experimental situation.
Here and henceforth, uniform means spatially constant.
In general, the laws need to be formulated locally, 
as described subsequently.

The classical zeroth law states that 
temperature $\vartheta$ is constant in thermal equilibrium.
The classical third law, expressed in its loosest form, is that 
absolute zero temperature cannot be physically attained. 
Thus $\vartheta>0$ in practice.

The entropy may be divided into entropy supply $\S$, given by
$$\vartheta\dot\S=\dot Q\eqno(2.4)$$
and entropy production $S-\S$. 
Then the second law may be written $$\dot S\ge\dot\S\eqno(2.5)$$
which expresses entropy production.
In words, $S$ is the entropy of the system,
where system means a comoving volume of material,
and $\S$ is the entropy supplied to the system.
Thus the second law implies that 
the total entropy of the universe cannot decrease.

Equality in the second law, 
$$\dot S=\dot\S\eqno(2.6)$$
holds in thermostatics, 
often called equilibrium thermodynamics or reversible thermodynamics. 
In the thermostatic case, the first and second laws for an inviscid fluid imply
$$\vartheta\dot S=\dot H+p\dot V\eqno(2.7)$$
which is the Gibbs equation.

It was found experimentally that the thermostatic state of a gas
could be described by just two quantities, 
such as density $\rho$ and temperature.
For uniform density, one may use the material volume $V$ instead of density,
since mass $M=V\rho$ is conserved:
$$\dot M=0.\eqno(2.8)$$
In particular, $H$ and $p$ are said to be state functions,
meaning functions of $(V,\vartheta)$.
If $S$ is also a state function, then the Gibbs equation can be rewritten 
in terms of the differential $\d$ in state space as
$\vartheta\d S=\d H+p\d V$, as Gibbs originally did [10].
This follows because the state functions $f$ are now uniform by assumption,
so that $\d f=\dot f\d t$.
However, in the non-uniform case, the last relation does not hold
and the two types of Gibbs equation cannot both be generally correct.
It turns out that the material Gibbs equation (2.7), localised,
forms part of the thermodynamic field equations [5], 
but that the state-space Gibbs equation is unnecessary.\footnote*
{For instance, in a static, spherically symmetric case with radius $r$,
one has $\d f=f'\d r$ and the state-space Gibbs equation implies
$\vartheta S'=H'+pV'$.
For liquid helium II at absolute zero,
this implies an unphysical equation of state $H=-pV$,
whereas the material Gibbs equation (2.7) would relate incompressibility,
$\dot V=0$, to conserved internal energy, $\dot H=0$.}
This point deserved emphasis because 
the Gibbs equation is often given in state-space form 
in introductory textbooks on thermodynamics, 
or on mechanics including thermodynamics,
as well as in much work on relativistic thermodynamics.
Indeed, this misunderstanding seems to be the root of 
the meaningless first and second laws mentioned above.
The thermodynamic theory developed in this article
will involve the material derivative rather than state-space differentials.
\bs\ni
{\bf III. Non-relativistic hydrodynamics}
\ms\ni
In non-relativistic hydrodynamics the basic quantities are 
the velocity vector $v$, density $\rho$ and stress tensor $\tau$ 
of the fluid.
The stress tensor is divided into thermostatic pressure $p$ 
and viscous stress $\sigma$ by
$$\tau=\sigma+ph\eqno(3.1)$$
where $h$ is the flat spatial metric.
Some authors use the opposite sign convention for stress,
but the above convention is standard in relativity.
The viscous stress $\sigma$ may be further divided into 
trace and traceless parts
$$\eqalignno
{&\varpi=\f13h:\sigma&(3.2a)\cr
&\varsigma=\sigma-\f13(h:\sigma)h&(3.2b)\cr}$$
where the colon denotes double symmetric contraction.
Then $\varpi$ is the viscous pressure (or bulk viscous stress)
and $\varsigma$ is the shear (or deviatoric) stress, satisfying
$$\sigma=\varsigma+\varpi h.\eqno(3.3)$$
The inviscid case is defined by $\sigma=0$.

Denoting the spatial volume form by $*$ 
and a spatial region by $\Sigma$, the volume is recovered as
$$V=\int_\Sigma{*}1\eqno(3.4)$$
and the mass as
$$M=\int_\Sigma{*}\rho.\eqno(3.5)$$
Note the kinematic relation
$$\dot{*}={*}D\cdot v\eqno(3.6)$$
where $D$ is the spatial gradient and the centred dot denotes contraction. 
Then conservation of mass (2.8)
may be written in a local form as the continuity equation
$$\dot\rho+\rho D\cdot v=0.\eqno(3.7)$$
Similarly, conservation of momentum, or Newton's first law for the fluid, 
yields the Euler-Cauchy equation
$$\rho(\dot v+D\Phi)=-D\cdot\tau\eqno(3.8)$$
where $\Phi$ is the Newtonian gravitational potential.
Conservation of mass and momentum are the basic hydrodynamic field equations.
\bs\ni
{\bf IV. Non-relativistic thermodynamics}
\ms\ni
The classical theory of thermodynamics must be reformulated locally
to be compatible with hydrodynamics.
This can be done by introducing local quantities 
which replace the classical integral quantities.
Firstly there are the internal energy density $\varepsilon$
and entropy density $s$, in terms of which
$$\eqalignno
{&H=\int_\Sigma{*}\varepsilon&(4.1a)\cr
&S=\int_\Sigma{*}s.&(4.1b)\cr}$$
Actually it is more traditional to use the specific internal energy 
$\epsilon_*=\varepsilon/\rho$
and the specific entropy $s_*=s/\rho$, 
due to frequent occurrences of combinations of the form
$$\rho\dot f_*=\dot f+f D\cdot v\eqno(4.2)$$
where $f_*=f/\rho$ for a function $f$;  
the identity follows from conservation of mass (3.7).

Heat supply is replaced locally by a heat flux vector $q$, in terms of which
$$\dot Q=-\oint_{\p\Sigma}{}\cdot q=-\int_\Sigma{*}D\cdot q\eqno(4.3)$$
where the second expression follows from the Gauss divergence theorem.
Similarly, defining the entropy flux vector
$\varphi=q/\vartheta$
allows the entropy supply to be recovered as
$$\dot\S=-\oint_{\p\Sigma}{}\cdot\varphi=-\int_\Sigma{*}D\cdot\varphi.
\eqno(4.4)$$
This relates to $\dot Q$ as in the classical definition (2.4)
if the temperature is uniform, $D\vartheta=0$.

The local first law is
$$\dot\varepsilon+\varepsilon D\cdot v=-D\cdot q-\tau:(D\otimes v)\eqno(4.5)$$
where $\otimes$ denotes the symmetric tensor product.
This local first law integrates to the classical form (2.2) 
for an inviscid fluid with uniform pressure, $Dp=0$.
Similarly, the local second law is
$$\dot s+s D\cdot v+D\cdot\varphi\ge0\eqno(4.6)$$
which integrates to the classical second law (2.5).
Lastly, the local form of the Gibbs equation is
$$\vartheta\dot s_*=\dot\epsilon_*-p\dot\rho/\rho^2.\eqno(4.7)$$
For uniform pressure and temperature, 
this integrates to the classical Gibbs equation (2.7).

It was found experimentally that the thermostatic state of a gas 
could be described by the density and temperature only.
Often this is generalised by giving two equations relating 
$(\rho,\vartheta,\epsilon,p)$ which specify a two-dimensional subspace.
However, this article will maintain $(\rho,\vartheta)$ 
as the state variables for definiteness.
In particular, $p$ and $\epsilon_*$ are functions of $(\rho,\vartheta)$,
these relations being called, respectively, 
thermal and caloric equations of state.
The simplest equations of state are those of an ideal gas: 
$$\eqalignno{&p=R_0\rho\vartheta&(4.8a)\cr
&\epsilon_*=c_0\vartheta&(4.8b)\cr}$$
where $R_0$ is the gas constant 
and $c_0$ is the specific heat capacity at constant density,
usually called constant volume.
The local Gibbs equation then integrates to give the specific entropy,
up to a time-independent function, as
$$s_*=c_0\log\vartheta-R_0\log\rho.\eqno(4.9)$$
In the inviscid case, the first law yields
$$-D\cdot q=c_0\rho\dot\vartheta-R_0\vartheta\dot\rho.\eqno(4.10)$$ 
Comparing with the classical definition of heat capacity $\dot Q/\dot\vartheta$,
implicitly assuming uniform temperature,
it can be seen that the specific heat capacities 
at constant density and pressure are $c_0$ and $c_0+R_0$ respectively.
Experimentally, $R_0$ and $c_0$ are nearly constant at low pressure, 
for a wide range of density and temperature [11].
Moreover, it is found that there is a universal gas constant 
$R_0m_0$ independent of the type of gas, where $m_0$ is the molecular mass,
and that monatomic gases satisfy 
$$c_0=\f32R_0.\eqno(4.11)$$ 
This relation and the ideal gas laws (4.8) 
can be derived in the kinetic theory of gases [5],
with $R_0m_0$ being the Boltzmann constant.

Completing the system of equations requires further equations for $(q,\sigma)$
which are consistent with the second law.
This is a tight restriction, as may be seen by eliminating 
the internal energy between the first law (4.5) and Gibbs equation (4.7), 
yielding
$$\dot s+s D\cdot v+D\cdot\varphi=
-{q\cdot D\vartheta\over{\vartheta^2}}
-{\sigma:(D\otimes v)\over{\vartheta}}.\eqno(4.12)$$
The simplest way to ensure compliance with the second law (4.6)
is for the right-hand side to be a sum of squares, which leads to 
$$\eqalignno{&q=-\kappa_0 D\vartheta&(4.13a)\cr
&\varpi=-\lambda_0D\cdot v&(4.13b)\cr
&\varsigma=-2\mu_0(D\otimes v-\f13(D\cdot v)h)&(4.13c)\cr}$$
with
$$\kappa_0\ge0\qquad\lambda_0\ge0\qquad\mu_0\ge0.\eqno(4.14)$$
If these coefficients are non-zero, 
the entropy production is given explicitly by
$$\dot s+s D\cdot v+D\cdot\varphi
={q\cdot q\over{\kappa_0\vartheta^2}}
+{\varpi^2\over{\lambda_0\vartheta}}
+{\varsigma:\varsigma\over{2\mu_0\vartheta}}\ge0\eqno(4.15)$$
which is manifestly non-negative.
This makes it clear that $(q,\sigma)$ cause entropy production,
i.e.\ they are thermally dissipative in nature.
In this article, $(q,\sigma)$ will be called the {\sl dissipative fields}
and the equations (4.13) determining them the {\sl dissipative relations}.
The dissipative relations and the equations of state 
are usually described collectively as the constitutive relations.

One may recognise the dissipative relation for $q$ as the Fourier equation 
and the dissipative relations for $\sigma$ 
as the definition of a Newtonian fluid;
substituting the latter into conservation of momentum (3.8) yields
$$\rho(\dot v+D\Phi)=-Dp+\mu_0D^2v+(\lambda_0+\f13\mu_0)D(D\cdot v)\eqno(4.16)$$
which is the Navier-Stokes equation.
These are the classical equations describing heat conduction 
and viscosity respectively, 
with $\kappa_0$ being the thermal conductivity,
$\mu_0$ the shear viscosity and $\lambda_0$ the bulk viscosity.
Originally, these dissipative relations were discovered experimentally,
with $\lambda_0=0$, but the above method shows that 
they may be loosely derived from the local second law.
This seems to have been originally noticed by Eckart [12],
who immediately generalised it to relativity [2].

Note that one of the dynamical equations is the local Gibbs equation (4.7),
even though this was originally concerned only with thermostatics.
Thus it has been implicitly assumed that,
in the general thermodynamic case, 
the Gibbs equation, as well as the thermal and caloric equations of state,
continue to take their thermostatic forms.
Also, note that there is no need for a state-space Gibbs equation
$\vartheta\d s_*=\d\epsilon_*+p\d(1/\rho)$;
indeed it would be generally inconsistent with 
the material Gibbs equation (4.7).
In this article, 
the material Gibbs equation and the relation $\varphi=q/\vartheta$
will be called {\sl entropic relations},
since they effectively define the entropic fields $(s,\varphi)$,
up to the usual ambiguity in $s$.
In the kinetic theory of gases, 
the entropic fields are defined in terms of the molecular distribution, 
allowing the entropic relations to be derived [5].

In summary, the basic thermohydrodynamic fields are
$(\rho,v,p,\sigma,\varepsilon,q,\vartheta,s,\varphi)$,
which constitutes twenty functions.
There are five conservation equations,
namely the first law and conservation of mass and momentum.
(The first law may be regarded as conservation of heat,
as will become clear in the relativistic theory).
The system is completed by two equations of state for $(p,\epsilon_*)$,
four components of the entropic relations for $(s,\varphi)$ 
and nine dissipative relations for $(q,\sigma)$,
the latter being chosen to imply the second law.

The fields pair up neatly with the equations expressing them 
or their material derivatives 
in terms of the basic fields and their spatial derivatives,
except for temperature and the first law;
even this becomes a temperature propagation equation 
on substituting other relations.
For instance, for an inviscid ideal gas at constant density $\rho=\rho_0$,
the local first law reads
$$c_0\rho_0\dot\vartheta=\kappa_0D^2\vartheta\eqno(4.17)$$
which is the diffusion equation.
This describes infinitely fast propagation of temperature
and is therefore physically implausible.
There is a similar problem of infinitely fast propagation 
implied by the Navier-Stokes equations [13].
Thus the standard local thermodynamics described so far,
e.g.\ by de~Groot \& Mazur [5], needs to be modified.
\bs\ni
{\bf V. Extended thermodynamics}
\ms\ni
The problem of infinite propagation speeds stems from the assumption that 
the entropy density $s$ satisfies the Gibbs equation (4.7),
originally intended to apply only to the thermostatic case.
A generalisation due to M\"uller [3]
involves allowing the the entropy to depend on 
quadratic terms in the dissipative fields $(q,\sigma)$.
This reduces to the foregoing theory 
in an approximation where the quadratic terms may be neglected,
i.e.\ the new theory is a quadratic rather than linear approximation 
to non-equilibrium.
Applying the second law as before yields modified dissipative relations 
containing relaxation terms that provide finite propagation speeds.
Moreover, dissipative relations of exactly this type 
may be obtained in the kinetic theory of gases 
by taking sufficiently many moments of the Boltzmann equation,
as originally shown by Grad [8].
M\"uller \& Ruggeri [13] now describe this as 
extended thermodynamics of irreversible processes,
reserving extended thermodynamics to describe 
a more systematic theory which assumes a further divergence-type balance law.

Since a distinction between entropy and thermostatic entropy is now required,
it is convenient to use $s_*$ for the thermostatic specific entropy henceforth,
with $s$ being the entropy density as before.
The equations displayed in the preceding subsection 
have been written with this distinction in mind,
e.g.\ the second law (4.6) contains $s$ 
and the Gibbs equation (4.7) contains $s_*$.
The thermostatic relation $s=\rho s_*$ is now modified 
to include quadratic terms in $(q,\sigma)$, giving the general form
$$s=\rho s_*
-\f12\rho(b_qq\cdot q+b_\varpi\varpi^2+b_\varsigma\varsigma:\varsigma).
\eqno(5.1)$$
Here the coefficients $b$ are arbitrary apart from being non-negative,
since entropy must reach a maximum in the thermostatic case:
$$b_{\{q,\varpi,\varsigma\}}\ge0.\eqno(5.2)$$
Similarly, the entropy flux $\varphi$ may differ from 
the thermostatic entropy flux $q/\vartheta$ by quadratic terms in $(q,\sigma)$,
giving the general form
$$\varphi={q\over{\vartheta}}-k_\varpi\varpi q-k_\varsigma\varsigma\cdot q.
\eqno(5.3)$$
Then the entropy production may be written explicitly as
$$\eqalignno{\dot s+s D\cdot v+D\cdot\varphi
&=-q\cdot\left({D\vartheta\over{\vartheta^2}}
+b_q\rho\dot q+k_\varsigma D\cdot\varsigma+k_\varpi D\varpi\right)\cr
&\quad-\varpi\left({D\cdot v\over{\vartheta}}
+b_\varpi\rho\dot\varpi+k_\varpi D\cdot q\right)\cr
&\quad-\varsigma:\left({D\otimes v\over{\vartheta}}
+b_\varsigma\rho\dot\varsigma+k_\varsigma D\otimes q\right)&(5.4)\cr}$$
where the coefficients $(b,k)$ have been assumed constant 
for the sake of simplicity.
So the second law suggests modified dissipative relations
$$\eqalignno{&q=-\kappa_0(D\vartheta+b_q\rho\vartheta^2\dot q
+k_\varsigma\vartheta^2D\cdot\varsigma+k_\varpi\vartheta^2D\varpi)&(5.5a)\cr
&\varpi=-\lambda_0(D\cdot v+b_\varpi\rho\vartheta\dot\varpi
+k_\varpi\vartheta D\cdot q)&(5.5b)\cr
&\varsigma=-2\mu_0(D\otimes v-\f13(D\cdot v)h
+b_\varsigma\rho\vartheta\dot\varsigma
+k_\varsigma\vartheta(D\otimes q-\f13(D\cdot q)h))&(5.5c)\cr}$$
which yield the entropy production with the same quadratic form as (4.15).
The key point here is that the relaxation coefficients $b_f$
introduce the time derivatives of the dissipative fields 
$f=(q,\varpi,\varsigma)$ into the dissipative relations,
giving equations for $f+t_f\dot f$, where the relaxation timescales are
$$\eqalignno{&t_q=\kappa_0b_q\rho\vartheta^2&(5.6a)\cr
&t_\varpi=\lambda_0b_\varpi\rho\vartheta&(5.6b)\cr
&t_\varsigma=2\mu_0b_\varsigma\rho\vartheta.&(5.6c)\cr}$$
This leads to finite propagation speeds, at least for linear perturbations [13].
The temperature propagation equation 
for an inviscid ideal gas at constant density now becomes
$$c_0\rho_0(\dot\vartheta+t_q\ddot\vartheta)=\kappa_0D^2\vartheta\eqno(5.7)$$
for constant $t_q$.
This is the telegraph equation, with finite propagation speed 
$\sqrt{\kappa_0/c_0\rho_0t_q}$.

In summary, M\"uller [3] obtained a causal theory of thermodynamics 
simply by allowing the entropic fields $(s,\varphi)$ 
to depend on quadratic terms in the dissipative fields $(q,\sigma)$.
There has been no tampering with the conservation laws, the second law,
the equations of state or even the Gibbs equation,
provided the latter is taken in thermostatic form,
i.e.\ in terms of the thermostatic specific entropy $s_*$.
This is often explained misleadingly as a generalisation of the Gibbs equation;
in reality, the thermostatic Gibbs equation is still assumed
but the entropy density is no longer assumed to be purely thermostatic.
One might continue to suppose that the thermostatic $s_*$ is a state function,
but the entropy density $s$ can no longer be a state function,
even for an ideal gas, 
thereby toppling what many textbooks erect as a principle of thermodynamics.
\bs\ni
{\bf VI. General relativity}
\ms\ni
General relativity is based on a four-dimensional manifold, space-time,
with a symmetric bilinear form $g$, the metric.
Taking units such that the Newtonian gravitational constant 
and the speed of light are unity,
the Einstein equation is
$$G=8\pi T\eqno(6.1)$$
where $G$ is the Einstein tensor of $g$ 
and $T$ is the energy tensor of the matter,
which will be taken in their contravariant forms.
The energy tensor is more fully described as the energy-momentum-stress tensor,
due to the physical interpretation of its various projections in a given frame 
as the energy density, momentum density or energy flux, and stress.

The contracted Bianchi identity, a purely geometrical identity, reads
$$\nabla\cdot G=0\eqno(6.2)$$
where $\nabla$ is the covariant derivative operator of $g$.
Therefore
$$\nabla\cdot T=0\eqno(6.3)$$
which expresses energy-momentum conservation.
A material model is specified by giving the general form of $T$
in terms of the material fields,
supplemented by any required equations other than energy-momentum conservation.
This paper gives such a procedure as a definition of thermodynamic matter.
Any such procedure requires at least a first law of thermodynamics.
Physically, the first law is an energy conservation equation,
but it differs from conservation of energy 
in the sense of a time-component of energy-momentum conservation (6.3).
The difference turns out to be conservation of mass.
In the Newtonian limit, conservation of energy yields 
conservation of mass to leading order and the first law as the next correction.
Thus relativistic thermodynamics requires a concept of mass.
In a microscopic description, this would be just the mass of the particles.

In relativity, mass is described by an energy-momentum vector $J$
which will be called the {\sl material current}.
Its magnitude is the density $\rho$ 
and its direction is the velocity vector $u$, assumed timelike:
$$\eqalignno{&J=\rho u&(6.4a)\cr &u\cdot u^\flat=-1.&(6.4b)\cr}$$
Here the sign convention is that spatial metrics are positive definite
and $\flat$ denotes the covariant dual with respect to $g$ (index lowering).
Similarly, $\sharp$ will denote the contravariant dual (index raising).
These accents are included for book-keeping 
and are irrelevant for many purposes.
In relativistic kinetic theory of gases [7],
$J$ is the first moment of the molecular distribution,
with $T$ being the second moment.
Conservation of mass and energy-momentum may then be derived 
from the relativistic Boltzmann equation.

Conservation of mass is expressed simply by
$$\nabla\cdot J=0.\eqno(6.5)$$
Written explicitly,
$$0=\nabla\cdot J=\dot\rho+\rho\nabla\cdot u\eqno(6.6)$$
where the overdot henceforth denotes the covariant derivative along $u$, 
$\dot f=u\cdot\nabla f$, which is the relativistic material derivative, 
or comoving derivative along the flow.
This has a similar form to non-relativistic conservation of mass (3.7).

Mass is a local form of energy in relativity, 
so it has an energy tensor $T_M$ 
which will be called the {\sl material energy tensor}.
This takes the standard form
$$T_M=\rho u\otimes u\eqno(6.7)$$
so that the relation 
$$\nabla\cdot J=-u\cdot(\nabla\cdot T_M)^\flat\eqno(6.8)$$
shows that conservation of mass is consistent with 
energy-momentum conservation for this type of matter, known as dust.
\bs\ni
{\bf VII. Heat}
\ms\ni
The proposed theory of relativistic thermodynamics
is based on four main physical principles.
The first principle is that (i) {\sl heat is a local form of energy}.
In general relativity, 
heat is therefore described by an energy tensor $T_H$ 
which will be called the {\sl thermal energy tensor}.
{\sl Thermodynamic matter} will be taken to mean a form of matter for which 
mass and heat are the only local forms of energy:
$$T=T_M+T_H.\eqno(7.1)$$
Comparing with a microscopic description, $T_H$ is the effective energy tensor
produced by the random motion of particles around the average flow given by $u$,
as can be made precise in relativistic kinetic theory of gases [7].

The physical meaning of the various components of the thermal energy tensor 
are now determined:
$$\varepsilon=u\cdot T_H^\flat\cdot u\eqno(7.2)$$
is the {\sl thermal energy density},
$$q=-\bot(T_H\cdot u^\flat)\eqno(7.3)$$
is the {\sl thermal flux} and
$$\tau=\bot T_H\eqno(7.4)$$
is the {\sl thermal stress}, 
where $\bot$ denotes projection by the spatial metric
$$h^{-1}=g+u^\flat\otimes u^\flat\eqno(7.5)$$ 
orthogonal to the fluid flow.
Note that $h$ denotes the contravariant form and $h^{-1}$ the covariant form.
Therefore
$$T_H=\varepsilon u\otimes u+2u\otimes q+\tau.\eqno(7.6)$$
As the notation indicates, thermal energy density, flux and stress
will be identified respectively with the internal energy density, 
heat flux and stress of the non-relativistic theory.
Therefore the energy tensor of the thermodynamic matter is
$$T=(\rho+\varepsilon)u\otimes u+2u\otimes q+\tau.\eqno(7.7)$$
This suffices to determine 
the quasi-local first law of relativistic thermodynamics 
in spherical symmetry, as is shown in the accompanying paper [1]. 
Written in terms of the effective density $\varrho=\rho+\varepsilon$,
the energy tensor has the same form as that of Eckart [2],
who defined specific internal energy as $\varrho/\rho$ 
plus an undetermined constant.
This is consistent, 
since specific thermal energy is $\varepsilon/\rho$.

The above principle implies that 
$(\varepsilon,q,\tau)$ are different aspects of heat.
In particular, one may say that 
thermal energy density or heat density $\varepsilon$ represents heat at rest, 
and $q$ heat in motion, relative to the average flow.
This furnishes the common-sense explanation for such simple physical processes 
as the heating, insulation and cooling of a body,
namely that heat flows in, remains at rest, then flows out.
This intuitive explanation is often mocked in thermodynamics textbooks,
accompanied by the claim that heat flux is the only form of heat.
In relativity, 
thermal flux is just one projection of a thermal energy tensor.

Similarly, what is traditionally called internal energy is just thermal energy, 
or simply heat, as the term is used in everyday language.
Moreover, the ambiguous zero of internal energy is fixed for thermal energy.
Recall that the original meaning of internal energy derived from the fact that
the measurable energies, due to heat flux and work, 
were not conserved.
According to conservation of energy, there had to be some other form of energy,
internal to the system.
The experiments of Joule established that 
this internal energy satisfied simple laws for an ideal gas.
This stop-gap concept of internal energy differs intrinsically 
from that of thermal energy,
yet they turn out to agree.
That internal energy is really thermal energy should anyway be clear 
even from non-relativistic kinetic theory of gases [5], 
where internal energy is defined as 
the average kinetic energy of the particles in the centre-of-momentum frame.

The second physical principle is (ii) {\sl conservation of mass}\/ (6.5).
As energy-momentum conservation (6.3) is automatic in general relativity,
conservation of mass is equivalent by (6.8) to {\sl conservation of heat} 
$$u\cdot(\nabla\cdot T_H)^\flat=0.\eqno(7.8)$$
Written explicitly,
$$0=-u\cdot(\nabla\cdot T_H)^\flat
=\dot\varepsilon+\varepsilon\nabla\cdot u+\nabla\cdot q
+q\cdot\dot u^\flat+\tau:(\nabla\otimes u^\flat).\eqno(7.9)$$
This reduces to the non-relativistic first law (4.5) in the Newtonian limit,
since the term in the acceleration $\dot u$ disappears,
as shown in the next section.
Thus principle (ii) is equivalent to the relativistic {\sl first law}.
This reveals that the first law is 
conservation of energy minus conservation of mass, cf.\ Eckart [2].
One may say that 
the first law is conservation of heat for the thermodynamic matter.
Note that this could be modified in the presence of other matter,
which would also contribute to the total energy tensor $T$.

Several points perhaps need to be stressed.
Firstly, although kinetic theory of gases has been mentioned for comparison,
the above theory of heat is manifestly independent of such microscopic theories.
Secondly, conservation of mass is required by the first law of thermodynamics.
In the perspective of general relativity, 
this is the primary distinction between thermodynamic matter 
and any other energy tensor $T$ to be used in the Einstein equation.
Thirdly, conservation of mass involves a material current which determines a preferred time direction $u$, the material frame of Eckart.
It is only in this frame that thermal flux---meaning the energy-flux component 
of the thermal energy tensor---agrees with what is normally called heat flux,
even in the limit of non-relativistic thermodynamics.
Finally, the basic conservation equations (6.3) and (6.5) are standard 
in relativistic thermodynamics and date back to Eckart [2].
What seems to have been lacking is the physical interpretation 
in terms of mass and heat.
In particular,
conservation of heat should not be confused with the old caloric theory,
as discussed further in the Conclusion.
\bs\ni
{\bf VIII. Newtonian limit}
\ms\ni
The Newtonian limit may be described in terms of a vector $\lambda$
which plays the role of Newtonian time.
The function $\Phi$ defined by
$$\e^{2\Phi}=-\lambda\cdot\lambda^\flat\eqno(8.1)$$
turns out to reduce to the Newtonian gravitational potential.
Similarly, the component of $u$ orthogonal to $\lambda$
turns out to reduce to the Newtonian velocity:
$$v=u+\e^{-2\Phi}(u\cdot\lambda^\flat)\lambda.\eqno(8.2)$$
Inverting,
$$\lambda={\e^\Phi(u-v)\over{\sqrt{1+|v|^2}}}.\eqno(8.3)$$
Taking units of length, factors of the speed $c$ of light 
may be introduced by the formal replacements
$$\eqalignno{&h\mapsto h&(8.4a)\cr
&(u,\lambda,v)\mapsto c^{-1}(u,\lambda,v)&(8.4b)\cr
&(\Phi,\rho)\mapsto c^{-2}(\Phi,\rho)&(8.4c)\cr
&(\varepsilon,\tau)\mapsto c^{-4}(\varepsilon,\tau)&(8.4d)\cr
&q\mapsto c^{-5}q.&(8.4e)\cr}$$
These factors are determined by the desired physical interpretation
of the fields in Newtonian theory.
This implicitly includes consequences like
$D\mapsto D$ for the covariant derivative of $h^{-1}$ 
and $\dot f\mapsto c^{-1}\dot f$ (where $f\mapsto f$)
for the material derivative $\dot f=u\cdot\nabla f$.
These replacements will be assumed in any subsequent equation involving $c$.

For $\lambda$ to qualify as a candidate for Newtonian time, it should satisfy
$$\bot L_\lambda h=O(c^{-2})\eqno(8.5)$$
where $L_\lambda$ is the Lie derivative along $\lambda$.
This condition reflects the absoluteness of Newtonian space.
Then
$$\bot(\nabla\otimes u^\flat)=D\otimes v^\flat+O(c^{-2})\eqno(8.6)$$
and in particular
$$\nabla\cdot u=D\cdot v+O(c^{-2}).\eqno(8.7)$$
So conservation of mass (6.6) and heat (7.9) read
$$\eqalignno{&\dot\rho+\rho D\cdot v=O(c^{-2})&(8.8a)\cr
&\dot\varepsilon+\varepsilon D\cdot v+D\cdot q+\tau:(D\otimes v^\flat)
=O(c^{-2})&(8.8b)\cr}$$
which manifestly reduce to non-relativistic conservation of mass (3.7) 
and heat (4.5).

Conservation of momentum may be written explicitly as
$$0=\bot(\nabla\cdot T)=(\rho+\varepsilon)\bot\dot u+\bot(q\cdot\nabla)u
+q(\nabla\cdot u)+\bot\dot q+\bot(\nabla\cdot\tau).\eqno(8.9)$$
Inserting factors of $c$ as in (8.4),
all terms except the first and last disappear in the Newtonian limit. 
A longer argument, skipped here, 
confirms that this reduces to non-relativistic conservation of momentum (3.8).
Thus all the conservation equations of non-relativistic thermodynamics
have been generalised to general relativity,
as conservation of energy-momentum and heat or mass.
Similarly, 
it is straightforward to check the Newtonian limits of the second law,
entropic relations and constitutive relations yet to be given.
\bs\ni
{\bf IX. Thermostatics and thermodynamics}
\ms\ni
The next step, giving equations of state,
requires a distinction between thermostatic and dissipative quantities,
e.g.\ between thermostatic pressure and viscous pressure.
Thus the thermal energy tensor $T_H$ is divided into 
a thermostatic part $T_0$ and a dissipative part $T_1$ by
$$T_H=T_0+T_1\eqno(9.1)$$
with $T_1=0$ in the thermostatic case.
The two are to be distinguished as in the non-relativistic theory
by the type of equations determining them:
$T_0$ is to be given by equations of state
and $T_1$ by dissipative relations consistent with the second law, 
the latter being described in the next section.

A {\sl fluid} may be defined by $T_0$ being isotropic, 
therefore taking the general form
$$T_0=\epsilon u\otimes u+ph.\eqno(9.2)$$
Then $p$ is the {\sl thermostatic pressure} 
and $\epsilon$ the {\sl thermostatic energy density}.
This isotropic form is appropriate for a fluid but not a solid;
for instance, an elastic solid generally has anisotropic thermostatic stress,
i.e.\ a general tensor replacing $ph$ in $T_0$.

The desired thermal and caloric equations of state may then be combined 
into an equation of state for $T_0$ in terms of $(\rho,\vartheta)$.
For instance, an ideal gas may be defined by
$$T_0=(c_0u\otimes u+R_0h)\rho\vartheta\eqno(9.3)$$
which gives thermal and caloric equations of state with the same form 
as for the non-relativistic ideal gas (4.8).
Generally, the equations determining $T_0$ 
may be called the {\sl thermostatic relations}.

In the thermostatic case $T_1=0$, 
this recovers the relativistic model of a perfect fluid [14], 
consisting of the energy tensor
$$T_M+T_0=(\rho+\epsilon)u\otimes u+ph\eqno(9.4)$$
together with equations of state for $(\epsilon,p)$ as functions of $\rho$
and conservation of mass or heat (7.9), which yields
$$\dot\epsilon_*=p\dot\rho/\rho^2\eqno(9.5)$$
in terms of the specific thermostatic energy 
$$\epsilon_*=\epsilon/\rho.\eqno(9.6)$$
For an ideal gas, this gives $\vartheta=f_0\rho^{R_0/c_0}$ 
for a time-independent function $f_0$.
Thus
$$\eqalignno{&\epsilon=c_0f_0\rho^{1+R_0/c_0}&(9.7a)\cr
&p=R_0f_0\rho^{1+R_0/c_0}.&(9.7b)\cr}$$
These equations of state specify a thermostatic ideal gas.
Actually, it is more common to use the effective density 
$\varrho=\rho+\epsilon$ 
instead of $\rho$ and $\epsilon$, forget mass conservation 
and give a single equation of state for $p$ as a function of $\varrho$.
Indeed, 
it is common to ignore $\epsilon$ and replace $\rho$ with $\varrho$ in (9.7b),
giving, for constant $f_0$, the polytropic equation of state.
This is a reasonable approximation if $\epsilon\ll\rho$,
as in Newtonian theory,
but the distinction between $\rho$ and $\varrho$ is important in general.

Returning to the general thermodynamic case, 
defining the {\sl dissipative stress} or viscous stress
$$\sigma=\tau-ph\eqno(9.8)$$
and the {\sl dissipative energy density}
$$\eta=\varepsilon-\epsilon\eqno(9.9)$$
allows the dissipative energy tensor to be composed as
$$T_1=\eta u\otimes u+2u\otimes q+\sigma\eqno(9.10)$$
which is the general form of such a tensor.
It will turn out that $\eta$ vanishes in the linear approximation for entropy, 
allowing consistency with standard non-relativistic thermodynamics [5].
Nevertheless, $\eta$ will be retained on the grounds that,
as a matter of principle, thermal energy need not be purely thermostatic.
The same distinction between thermostatic and dissipative energy density
could be drawn in the non-relativistic case,
leaving ambiguities to be fixed in equations where $\eta$ might occur;
these were fixed in equations displayed in Section~IV by using 
$\varepsilon$ and $\epsilon_*$ with appropriate foresight.

Exhibiting all terms in the energy tensor,
$$T=T_M+T_0+T_1=(\rho+\epsilon+\eta)u\otimes u+2u\otimes q+ph+\sigma.
\eqno(9.11)$$
This shows that $T$ incorporates all the basic thermohydrodynamic fields 
of the non-relativistic theory described previously, 
apart from temperature $\vartheta$ and entropy $(s,\varphi)$,
but adding dissipative energy density $\eta$.
\bs\ni
{\bf X. Entropy}
\ms\ni
The third physical principle is that (iii) {\sl entropy is a local current}.
In relativity, it is therefore described by
an {\sl entropy current} vector $\Psi$ whose components are
{\sl entropy density}
$$s=-u^\flat\cdot\Psi\eqno(10.1)$$ 
and {\sl entropy flux}
$$\varphi=\bot\Psi.\eqno(10.2)$$
As their names indicate, entropy density and entropy flux 
will be identified with their non-relativistic versions.
Therefore
$$\Psi=su+\varphi.\eqno(10.3)$$
The fourth physical principle is (iv) {\sl non-destruction of entropy}.
In relativity, this is therefore expressed by
$$\nabla\cdot\Psi\ge0.\eqno(10.4)$$
Explicitly,
$$\dot s+s\nabla\cdot u+\nabla\cdot\varphi\ge0\eqno(10.5)$$
which reduces to the non-relativistic second law (4.6) in the Newtonian limit.
Thus principle (iv) is the relativistic {\sl second law}.
These two principles are standard in relativistic thermodynamics.
Again it seems to be Eckart [2] who first gave a version of (10.5),
assuming $\varphi=q/\vartheta$.

The final stage consists of introducing entropic relations for $\Psi$
in terms of the other fields,
then finding dissipative relations for $T_1$ which imply the second law.
Successive approximations for $\Psi$ will be taken:
$\Psi=\Psi_0$ for the thermostatic case,
$\Psi=\Psi_0+\Psi_1$ for the linear approximation 
and $\Psi=\Psi_0+\Psi_1+\Psi_2$ for the quadratic approximation for entropy.

The thermostatic entropy current $\Psi_0$ may be assumed to satisfy
$$\eqalignno{&\bot\Psi_0=0&(10.6a)\cr
&\nabla\cdot\Psi_0=-{u\over{\vartheta}}\cdot(\nabla\cdot T_0)^\flat.
&(10.6b)\cr}$$
The first equation states that 
the thermostatic entropy current is comoving with the material.
Therefore $\Psi_0$ may be written in terms of the specific thermostatic entropy 
$$s_*=-u^\flat\cdot\Psi_0/\rho.\eqno(10.7)$$ 
For a fluid (9.2), the second equation may be written as
$$\vartheta\nabla\cdot\Psi_0
=\dot\epsilon+\epsilon\nabla\cdot u+p\nabla\cdot u\eqno(10.8)$$
or
$$\vartheta\dot s_*=\dot\epsilon_*-p\dot\rho/\rho^2.\eqno(10.9)$$
This form was given by Eckart [2] and is identical to that of 
the non-relativistic local Gibbs equation (4.7).\footnote*
{Most references on relativistic thermodynamics 
give a different Gibbs equation.
Sometimes the material derivative is replaced with the covariant derivative 
in the Gibbs equation [15,16],
giving an equation whose spatial components are quite unnecessary 
and generally overdetermine the fields.
This equation has even been described as the first law.
In other cases, 
the material derivative is replaced with a state-space differential 
in the Gibbs equation [4,6],
which generally does not imply the material Gibbs equation (10.9).
It has even been claimed that the thermostatic $(\Psi_0,T_0)$
should be replaced with $(\Psi,T)$ in this state-space equation [4].}
This is usually regarded as defining thermostatic temperature.
For an ideal gas (9.3), it integrates to
$$s_*=c_0\log\vartheta-R_0\log\rho\eqno(10.10)$$
which has the same form as in the non-relativistic case (4.9).

The relations (10.6) therefore constitute a generalised Gibbs equation, 
relativistically unified in terms of $\Psi_0$ and $T_0$.
This relativistic Gibbs equation has the desired property that 
entropy is conserved in the thermostatic case:
$$T_1=0\quad\Rightarrow\quad\nabla\cdot\Psi_0=\rho\dot s_*=0\eqno(10.11)$$
as follows from conservation of heat (7.8).
In other words, thermostatics requires equality in the second law.
Indeed, one might propose a stronger version of the second law,
to the effect that non-thermostatic processes necessarily produce entropy.
One could then define thermostatics,
or equilibrium thermodynamics, or reversible thermodynamics,
by equality in the second law (10.4).
\bs\ni
{\bf XI. Linear approximation}
\ms\ni
The linear correction $\Psi_1$ to entropy current may be assumed as
$$\Psi_1=-{u^\flat\over{\vartheta}}\cdot T_1\eqno(11.1)$$
which may be regarded as defining temperature $\vartheta$ away from equilibrium.
Then the entropy flux $\bot\Psi_1=q/\vartheta$ has the same form 
as in non-relativistic thermodynamics.
The component of $\Psi_1$ along the flow, $\eta/\vartheta$, 
is not so determined, except that the above linear form is the simplest.
However, it turns out to yield a useful cancellation in the entropy production, 
which, using conservation of heat (7.8), becomes simply
$$\nabla\cdot(\Psi_0+\Psi_1)
=-T_1:\left(\nabla\otimes{u^\flat\over{\vartheta}}\right).\eqno(11.2)$$
Thus the entropy production is a contraction 
of the dissipative energy tensor $T_1$ 
with the gradient of the inverse temperature vector $u/\vartheta$,
thereby unifying what are sometimes called, respectively,
the thermodynamic fluxes and thermodynamic forces [5].
Writing the components explicitly,
$$\nabla\cdot(\Psi_0+\Psi_1)
=-{q\cdot(\nabla\vartheta+\vartheta\dot u^\flat)\over{\vartheta^2}}
-{\sigma:(\nabla\otimes u^\flat)\over{\vartheta}}
-{\eta\dot\vartheta\over{\vartheta^2}}.\eqno(11.3)$$
So the second law (10.4) suggests the dissipative relations
$$\eqalignno{&q=-\kappa_0\bot(\nabla^\sharp\vartheta+\vartheta\dot u)&(11.4a)\cr
&\varpi=-\lambda_0\nabla\cdot u&(11.4b)\cr
&\varsigma=-2\mu_0\bot(\nabla^\sharp\otimes u-\f13(\nabla\cdot u)h)&(11.4c)\cr
&\eta=-\nu_0\dot\vartheta&(11.4d)\cr}$$
with
$$\kappa_0\ge0\qquad\lambda_0\ge0\qquad\mu_0\ge0\qquad\nu_0\ge0.\eqno(11.5)$$
If these coefficients are non-zero, 
the entropy production is given explicitly by
$$\nabla\cdot\Psi={q\cdot q^\flat\over{\kappa_0\vartheta^2}}
+{\varpi^2\over{\lambda_0\vartheta}}
+{\varsigma:\varsigma^\flat\over{2\mu_0\vartheta}}
+{\eta^2\over{\nu_0\vartheta^2}}\ge0.\eqno(11.6)$$
The dissipative relation for $q$ is a relativistic modification 
of the Fourier equation, including a term in the acceleration $\dot u$ 
which disappears in the Newtonian limit.
The dissipative relations for $\sigma$ are relativistic versions of 
the viscosity relations for a Newtonian fluid, 
yielding a relativistic Navier-Stokes equation.
With $\varpi=0$, these were originally derived by Eckart [2].
The last dissipative relation for $\eta$ appears to be new
and modifies the temperature propagation equation: 
taking the example of an inviscid ideal gas at constant density 
with vanishing acceleration,
the first law (7.9) becomes
$$\rho_0c_0\dot\vartheta-\nu_0\ddot\vartheta=\kappa_0D^2\vartheta.\eqno(11.7)$$
Thus the parabolic (diffusion) equation for $\nu_0=0$ becomes  
a hyperbolic (telegraph) equation for $\nu_0<0$, as desired,
but an elliptic equation for $\nu_0>0$.
Since ellipticity is even worse than parabolicity on causal grounds,
this leads to $\nu_0=0$ and therefore $\eta=0$ as promised.
Thus the linear approximation for entropy recovers and generalises 
the Eckart theory [2], 
which concerned the case of a fluid without bulk viscosity.
\bs\ni
{\bf XII. Quadratic approximation}
\ms\ni
For the quadratic correction $\Psi_2$ to the entropy current,
the most general form which is quadratic 
in the dissipative energy tensor $T_1$ is
$$\Psi_2=-\f12(b_qq\cdot q^\flat+b_\varpi\varpi^2
+b_\varsigma\varsigma:\varsigma^\flat+b_\eta\eta^2)\rho u
-k_\varpi\varpi q-k_\varsigma\varsigma\cdot q^\flat-k_\eta\eta q\eqno(12.1)$$
where
$$b_{\{q,\varpi,\varsigma,\eta\}}\ge0\eqno(12.2)$$
so that entropy is maximised in equilibrium.
The coefficients $b$ lead to relaxation effects as in the non-relativistic case.
Explicitly, the entropy production is
$$\eqalignno{\nabla\cdot(\Psi_0+\Psi_1+\Psi_2)
&=-q\cdot\left({\nabla\vartheta+\vartheta \dot u^\flat\over{\vartheta^2}}
+b_q\rho\dot q^\flat
+k_\varsigma(\nabla\cdot\varsigma)^\flat
+k_\varpi\nabla\varpi
+k_\eta\nabla\eta\right)\cr
&\quad-\varpi\left({\nabla\cdot u\over{\vartheta}}
+b_\varpi\rho\dot\varpi
+k_\varpi\nabla\cdot q\right)\cr
&\quad-\varsigma:\left({\nabla\otimes u^\flat\over{\vartheta}}
+b_\varsigma\rho\dot\varsigma^\flat
+k_\varsigma\nabla\otimes q^\flat\right)\cr
&\quad-\eta\left({\dot\vartheta\over{\vartheta^2}}
+b_\eta\rho\dot\eta
+k_\eta\nabla\cdot q\right)&(12.3)\cr}$$
where the coefficients $(b,k)$ have been assumed constant 
for the sake of simplicity,
the generalisation being straightforward.
The entropy production takes the same quadratic form (11.6) 
as in the linear approximation 
if the dissipative relations are taken as
$$\eqalignno{&q=-\kappa_0\bot(\nabla^\sharp\vartheta+\vartheta\dot u
+b_q\rho\vartheta^2\dot q
+k_\varsigma\vartheta^2\nabla\cdot\varsigma
+k_\varpi\vartheta^2\nabla^\sharp\varpi
+k_\eta\vartheta^2\nabla^\sharp\eta)&(12.4a)\cr
&\varpi=-\lambda_0(\nabla\cdot u
+b_\varpi\rho\vartheta\dot\varpi
+k_\varpi\vartheta\nabla\cdot q)&(12.4b)\cr
&\varsigma=-2\mu_0\bot(\nabla^\sharp\otimes u-\f13(\nabla\cdot u)h
+b_\varsigma\rho\vartheta\dot\varsigma
+k_\varsigma\vartheta(\nabla^\sharp\otimes q-\f13h:(\nabla\otimes q^\flat)h))
&(12.4c)\cr
&\eta=-\nu_0(\dot\vartheta
+b_\eta\rho\vartheta^2\dot\eta
+k_\eta\vartheta^2\nabla\cdot q).&(12.4d)\cr}$$
This completes the system of thermodynamic field equations
for specified coefficients $(b,k)$.
In the case $\eta=0$, 
these dissipative relations reduce to those of Israel \& Stewart [4],
also given by Jou et al.\ [17].
If the coefficients $(b,k)$ are not constant, additional derivatives appear,
as noted by Hiscock \& Lindblom [18].

Relativistic kinetic theory of gases [7] 
confirms the existence of dissipative energy density $\eta$;
indeed it emerged unrecognised in the non-relativistic theory of Grad [8]
as essentially the 14th moment of the molecular distribution,
the previous 13 moments being $(\rho,v,\epsilon,\varsigma,q)$,
with $(\vartheta,p)$ being given by equations of state and $\varpi$ vanishing.
For instance, for an monatomic (4.11) ideal (9.3) gas with $\varpi=0$, setting
$$b_q={2\over{5p^2\vartheta}}\qquad
k_\varsigma={2\over{5p\vartheta}}\qquad
b_\varsigma={1\over{2p\rho\vartheta}}\qquad
k_\eta={8\rho\over{15p^2\vartheta}}\qquad
b_\eta={8\rho\over{15p^3\vartheta}}\eqno(12.5)$$
and
$$\kappa_0={5p^2\over{2B_0\rho^2\vartheta}}\qquad
\mu_0={2p\over{3B_0\rho}}\qquad
\nu_0={15p^3\over{8B_0\rho^3\vartheta}}\eqno(12.6)$$
where $B_0$ is a constant,
the dissipative relations in the non-relativistic limit become
$$\eqalignno{&0=\dot q+B_0\rho q+\f52R_0^2\rho\vartheta D\vartheta+
R_0\vartheta D\cdot\varsigma+\f43D\eta&(12.7a)\cr
&0=\dot\varsigma+\f32B_0\rho\varsigma
+2R_0\rho\vartheta(D\otimes v-\f13(D\cdot v)h)
+\f45(D\otimes q-\f13(D\cdot q)h)&(12.7b)\cr
&0=\dot\eta+B_0\rho\eta+R_0\vartheta D\cdot q&(12.7c)\cr}$$
where the speed of light has been introduced as in (8.4),
with $R_0\mapsto c^2R_0$, $B_0\mapsto cB_0$,
$\vartheta\mapsto c^{-4}\vartheta$
and $\eta\mapsto c^{-6}\eta$.
Together with the non-relativistic conservation laws,
these are the 14-moment equations (1.42) of M\"uller \& Ruggeri [13],
their 14th moment being $8\eta$, 
with the factor determined in relativistic kinetic theory [7].
This confirms the existence of dissipative energy density
and thereby topples another pillar of textbook thermodynamics,
that internal energy necessarily be a state function.
\bs\ni
{\bf XIII. Integral quantities and laws}
\ms\ni
There exist integral forms of the second law and conservation of mass, 
because the local forms are current (semi-) conservation laws,
allowing the Gauss divergence theorem to be applied.
In particular, 
the mass of a region $\Sigma$ of a spatial hypersurface may be defined by
$$M(\Sigma)=-\int_\Sigma J\eqno(13.1)$$
where the volume form and contraction with unit normal are implicit.
Consider a space-time region $\Omega$ bounded by $\Sigma$, a later $\Sigma'$ 
and a hypersurface region $\hat\Sigma$ generated by flowlines of $u$.
Applying the Gauss theorem,
$$M(\Sigma')-M(\Sigma)=\int_\Omega\nabla\cdot J=0\eqno(13.2)$$
which is the integral form of conservation of mass.
Similarly, the entropy of $\Sigma$ may be defined by
$$S(\Sigma)=-\int_\Sigma\Psi\eqno(13.3)$$
and the entropy supply through $\hat\Sigma$ by
$$\S(\hat\Sigma)=-\int_{\hat\Sigma}\Psi.\eqno(13.4)$$
Then the Gauss theorem yields
$$S(\Sigma')-S(\Sigma)-\S(\hat\Sigma)=\int_\Omega\nabla\cdot\Psi\ge0
\eqno(13.5)$$
which is the integral second law.
Thus there is an integral second law of thermodynamics 
even in general relativity.
An integral first law may seem awkward in similar generality, 
since the local first law (7.8) is not a current conservation law, 
but in spherical symmetry 
a simple integral (or quasi-local) first law exists,
as shown in the accompanying paper [1].

The integral laws become particularly simple if the vorticity vanishes,
$$\bot(\nabla\wedge u^\flat)=0\eqno(13.6)$$
where $\wedge$ denotes the antisymmetric tensor product.
Then there exist spatial hypersurfaces orthogonal to $u$ 
which may be used to define preferred volume integrals.
For instance, the mass is
$$M=\int_\Sigma{*}\rho\eqno(13.7)$$
where $\Sigma$ henceforth denotes a region of a hypersurface orthogonal to $u$, 
with volume form $*$.
The induced metric of $\Sigma$ is just $h^{-1}$, which defines $*$.
One may assume that $\Sigma$ has finite volume,
or infinite volume with all relevant integrals existing.

Concerning time derivatives:
in the relativistic case, $\dot M$ is generally meaningless,
since proper time along $u$ generally does not label 
the hypersurfaces orthogonal to $u$.
Labelling these hypersurfaces by $t$, 
choosing vanishing shift vector 
and denoting $\Delta f=\p f/\p t$ for functions $f$, 
the relationship
$$\dot f=\e^{-\phi}\Delta f\eqno(13.8)$$
defines the function $\phi$, up to relabellings of $t$.
Then $\Delta M$ is meaningful and the Gauss theorem yields
$$\int_\Omega\nabla\cdot J=\int_{\Sigma'}{*}\rho-\int_\Sigma{*}\rho
\eqno(13.9)$$
which differentiates to
$$\int_\Sigma{*}\e^\phi\nabla\cdot J=\Delta M.\eqno(13.10)$$
Therefore local conservation of mass (7.10) implies the integral form
$$\Delta M=0\eqno(13.11)$$
which has the same form as classical conservation of mass (2.8).

Similarly, the entropy is 
$$S=\int_\Sigma{*}s\eqno(13.12)$$
and the entropy supply through $\hat\Sigma$ is
$$\S=-\int_{\hat\Sigma}{\hat*}\cdot\varphi\eqno(13.13)$$
where $\hat*$ is the volume form of $\hat\Sigma$ 
times the unit outward normal covector tangent to $\Sigma$.
Then
$$\Delta\S=-\oint_{\p\Sigma}{}\cdot\e^\phi\varphi\eqno(13.14)$$
where the surface $\p\Sigma$ is the intersection of $\Sigma$ and $\hat\Sigma$,
with area form and unit normal implicit in $\oint$.
This has the same form as the classical definition of entropy supply (4.4),
apart from the use of $\Delta$ and the corresponding factor $\e^\phi$,
which tends to one in the Newtonian limit.
Applying the Gauss theorem again,
$$\int_\Omega\nabla\cdot\Psi=\int_{\Sigma'}{*}s-\int_\Sigma{*}s
+\int_{\hat\Sigma}\hat{*}\cdot\varphi\eqno(13.15)$$
which differentiates to
$$\int_\Sigma{*}\e^\phi\nabla\cdot\Psi=\Delta S-\Delta\S.\eqno(13.16)$$
Therefore $\nabla\cdot\Psi$ is the entropy density production rate 
and the local second law (10.4) integrates to
$$\Delta S\ge\Delta\S\eqno(13.17)$$
which has the same form as the classical second law (2.5).

Similarly,
the heat supply through $\hat\Sigma$ may be defined by
$$Q=-\int_{\hat\Sigma}{\hat*}\cdot q\eqno(13.18)$$
which yields
$$\Delta Q=-\oint_{\p\Sigma}{}\cdot\e^\phi q\eqno(13.19)$$
again with the same form as the classical definition of heat supply (4.3).
Finally, the thermal energy may be defined by
$$H=\int_\Sigma{*}\varepsilon.\eqno(13.20)$$
The forms 
$$\eqalignno{&M=\int_\Sigma{*}u\cdot T_M^\flat\cdot u&(13.21a)\cr
&H=\int_\Sigma{*}u\cdot T_H^\flat\cdot u&(13.21b)\cr}$$ 
explain the notation adopted earlier:
the material and thermal energy tensors $T_M$ and $T_H$ 
are the energy tensors of mass and heat respectively.

Thermal energy $H$ is what thermodynamicists call internal energy
and what ordinary people call heat.
In particular, this defines the heat of a body or volume of fluid.
Thermodynamics textbooks often claim that there is no such thing as 
the heat of a body, in bizarre conflict with common experience.
This seems to involve the belief that
heat flux is the only form of heat,
i.e.\ what was really meant is that
there is no conserved quantity corresponding to the flux of heat,
as supposed by the old caloric theory.
A nice analogy [11], 
intended to support the above claim but actually revealing the opposite,
is that there is no such thing as the amount of rain in a lake.
True, because rain is not the only form of water,
but the amount of water in the lake does make sense.
\bs\ni
{\bf XIV. Conclusion}
\ms\ni
A general macroscopic theory of generally relativistic thermodynamics 
has been developed,
based on clear principles and sufficient to include relaxation effects.
There is a full set of dynamical equations, easily applied in practice,
with no meaningless derivatives or other common thermostatic confusions.
The basic fields are the temperature $\vartheta$,
the material, thermostatic and dissipative energy tensors $(T_M,T_0,T_1)$ 
and the entropy current $\Psi$.
These fields may be paired respectively with the thermodynamic field equations:
conservation of mass (6.5) or heat (7.8),
conservation of energy-momentum (6.3),
the thermostatic relations for $T_0$,
e.g.\ the equations of state (9.3) for an ideal gas,
the dissipative relations for $T_1$
and the entropic relations for $\Psi$:
thermostatic (10.6), linear (11.1) and quadratic (12.1) approximations.
The dissipative relations (11.4) or (12.4) 
are chosen to yield the second law (10.4), subject to relativistic causality.
Having ensured these conditions, the dynamical system may be reduced 
by eliminating $\Psi$ and the entropic relations,
since $\Psi$ does not appear in the remaining equations.
Similarly, the thermostatic relations may be used to eliminate further fields,
numbering two in the case of a fluid.

Four physical principles have been emphasised, which may be regarded as 
minimal requirements for a definition of thermodynamic matter.
In principle this includes solids, necessarily imperfect in relativity,
though the form of the thermostatic relations 
has been specified only for fluids.
Three of the principles are standard.
However, the first principle, that heat is a local form of energy,
seems not to have been clearly stated before.
This principle unifies various thermodynamic quantities 
in an essentially relativistic way 
as the energy, flux and stress components of a thermal energy tensor.
This unification is analogous to that of electricity and magnetism
in relativistic electromagnetism.
Moreover, this agrees with non-relativistic thermodynamics
in the physical interpretation of thermal flux only in the material frame.
The principle thereby resolves the controversy 
of Eckart [2] versus Landau-Lifshitz [6] frame.

A further consequence of this formulation of basic principles is that 
it is unnatural to insist that thermal energy be purely thermostatic,
despite the traditional dogma that internal energy be a state function.
The similarly traditional dogma that entropy be a state function
had already disappeared in M\"uller's extended thermodynamics [3].
Thus we have an apparently new thermodynamical quantity, dissipative energy,
which nevertheless occurred unrecognised in the kinetic theory of gases [7].
These former assumptions may now be derived 
in the linear approximation to non-equilibrium, 
but need not hold in general, e.g.\ in the quadratic approximation.

One artificial feature of this theory,
shared with the original M\"uller [3] and Israel-Stewart [4] theories,
is that the quadratic correction to entropy current 
involves certain coefficients which are left unspecified.
However, these coefficients can be determined for an ideal gas 
by kinetic theory of gases [13].
Kinetic theory also automatically implies 
conservation of mass and energy-momentum, 
and higher moments can be used to obtain a hierarchy of divergence equations 
for which there is a standard mathematical theory 
of local existence, uniqueness and causality [19].
However, this type of relativistic extended thermodynamics
has so far been worked out 
under the assumption of vanishing dissipative energy density [13].
The main physical problem with this type of theory is that 
it is not known how to directly measure the higher moments.

It seems necessary to stress that 
conservation of heat for thermodynamic matter has been derived 
from conservation of mass and energy-momentum, 
and definitely differs from the old caloric theory of Carnot [20].
A relativistic version of the caloric theory would be 
to describe heat by an energy-momentum vector, 
the caloric current $J_H=\varepsilon u+q$,
and conservation of heat by $\nabla\cdot J_H=0$,
as for conservation of mass.
This would integrate by (13.18) and (13.20) to $\Delta H=\Delta Q$.
Carnot's theory of heat can be interpreted as involving this equation 
and the claim that the heat $H$ is a state function, 
which contradicts experiment.
Unfortunately the discrepancy between this and the classical first law (2.1) 
has led to a historical over-reaction against the intuitive notion of heat,
specifically the widespread insistence that 
the internal energy of Clausius differs from the heat of Carnot,
whereas Carnot's mistake was instead in the formulation of conservation of heat.
The relativistic theory reveals that the caloric theory failed only because 
the mathematical nature of energy---an energy-momentum-stress tensor $T_H$ 
rather than an energy-momentum vector $J_H$---was not understood 
in pre-relativistic physics.
In particular, 
material stress appears in the form of work in the first law, 
because it is also a form of heat.
If the thermodynamic matter is not isolated,
heat and other forms of energy may be exchanged.

The relativistic theory similarly resolves the ancient philosophical question 
of whether heat is a substance.
Put simply, both mass and heat are forms of energy, but with different natures:
mass is described by a current $J$ 
which satisfies a vectorial conservation law (6.5), conservation of mass,
whereas heat is a general form of energy $T_H$ 
satisfying a tensorial conservation law (7.8), the first law of thermodynamics.
\np\ni
{\bf Acknowledgements}\par\ni
Research supported by a European Union Science and Technology Fellowship,
formerly by a Japan Society for the Promotion of Science 
postdoctoral fellowship.
Part of the work was carried out during a visit 
to the Albert Einstein Institute in Potsdam.
Thanks are due to J\"urghen Ehlers for the latter hospitality 
and for tracking down an error in an early version of the theory,
and to Ingo M\"uller for explaining various aspects of extended thermodynamics.
\bs
\begingroup 
\parindent=0pt\everypar={\global\hangindent=20pt\hangafter=1}\par
{\bf References}\ms
\ref{[1] Hayward S A 1998}\CQG{15}{3147}
\ref{[2] Eckart C 1940}\PR{58}{919}
\ref{[3] M\"uller I 1967}\ZP{198}{329}
\ref{[4] Israel W \& Stewart J M 1979}\AP{118}{341}
\refb{[5] de Groot S R \& Mazur P 1962}{Non-equilibrium Thermodynamics}
{(North-Holland)}
\refb{[6] Landau L \& Lifshitz E M 1987}{Fluid Mechanics}{(Addison-Wesley)}
\refb{[7] Hayward S A}{Relativistic kinetic theory of gases}
{(in preparation)} 
\ref{[8] Grad H 1949}\CPAM2{331}
\refb{[9] Truesdell C 1984}{Rational Thermodynamics}{(Springer-Verlag)}
\refb{[10] Gibbs J W 1960}{Elementary Principles in Statistical Mechanics}
{(Dover)}
\refb{[11] Zemansky M W 1968}{Heat and Thermodynamics}{(McGraw-Hill)}
\ref{[12] Eckart C 1940}\PR{58}{267}
\refb{[13] M\"uller I \& Ruggeri T 1993}{Extended Thermodynamics}
{(Springer-Verlag)}
\refb{[14] Schutz B F 1985}{A First Course in General Relativity}
{(Cambridge University Press)}
\refb{[15] Novikov I D \& Thorne K S 1973 in}{Black Holes}
{ed.~DeWitt C \& DeWitt B S (Gordon \& Breach)}
\refb{[16] Misner C W, Thorne K S \& Wheeler J A 1973}{Gravitation}{(Freeman)}
\refb{[17] Jou D, Casas-V\'azquez J \& Lebon G 1996}
{Extended Irreversible Thermodynamics}{(Springer-Verlag)}
\ref{[18] Hiscock W A \& Lindblom L 1983}\AP{151}{466}
\ref{[19] Geroch R \& Lindblom L 1990}\PR{D41}{1855}
\refb{[20] Carnot S}{Reflections on the motive power of fire}
{ed.\ Mendoza E 1960 (Dover)}
\endgroup
\bye